\documentclass[superscriptaddress,aps,prd,nofootinbib,twocolumn,showpacs]{revtex4-1}
\usepackage[english]{babel}
\usepackage[utf8]{inputenc}
\usepackage{float}
\usepackage{graphicx}
\usepackage[caption=false]{subfig}
\usepackage{color}
\usepackage{hyperref}
\usepackage{amsfonts}
\usepackage{amsmath}
\usepackage{amssymb}
\usepackage{graphicx}
\usepackage{mathrsfs}
\usepackage{multirow}

\def\be {\begin{equation}}
\def\ee {\end{equation}}
\def\bea {\begin{eqnarray}}
\def\eea {\end{eqnarray}}
\def\bc {\begin{center}}
\def\ec {\end{center}}

\begin{document}

\title{Unveiling the correlations of tidal deformability 
with  the nuclear symmetry energy parameters}

 \author{Tuhin Malik}
 \affiliation{BITS-Pilani, Department of Physics, Hyderabad Campus, Hyderabad - 500078, India}
  
 \author{B. K. Agrawal}
 \affiliation{Saha Institute of Nuclear physics, Kolkata 700064, India}
 \affiliation{Homi Bhabha National Institute, Anushakti Nagar, Mumbai -
 400094, India}
 \author{Constan\c ca Provid\^encia}
 \affiliation{CFisUC, Department of Physics,
 University of Coimbra, P-3004 - 516  Coimbra, Portugal}

 \author{J. N. De}
 \affiliation{Saha Institute of Nuclear physics, Kolkata 700064, India} 
 
\date{\today} 

\begin{abstract} 

The chi-squared  based covariance approach allows one to estimate the
correlations among desired observables related to nuclear matter directly
from a set of fit data without taking recourse to the distributions of
the nuclear matter parameters (NMPs). Such an approach is applied to
study the correlations of tidal deformability of neutron star with the
slope and the curvature parameters of  nuclear symmetry energy governed
by  an extensive set of fit data on the finite nuclei together with the
maximum mass of the neutron star. The knowledge of the distributions of
NMPs consistent with the fit data is implicitly inbuilt in the Hessian
matrix which is central to this covariance approach.  Comparing our
results with those obtained with the explicit use of the distributions
of NMPs, we show that the appropriate correlations among NMPs as induced
by the fit data are instrumental in strengthening the  correlations of
the tidal deformability with the symmetry energy parameters, without it,
the said correlations tend to disappear. The interplay between isoscalar and isovector NMPs is also emphasized.

\end{abstract}

\maketitle

{\it Introduction--} \label{intro}
The determination of the equation of state (EoS) of nuclear matter over
a large density range, much beyond the saturation density $\rho_0$ is
one of the main objectives of both nuclear physics and astrophysics to
date \cite{Haensel2007,Lattimer2015,Rezzolla2018}. The neutron stars
(NSs), believed to contain nuclear matter upto few times $\rho_0$ in
their core are the ideal cosmic laboratories to explore the nuclear EoS,
complemented with observations from terrestrial experiments. To understand
the internal structure of the NS and its properties such as its crust,
mass, radius, quadrupole deformation, moment of inertia etc, one  needs
to connect different branches of physics that include low energy nuclear
physics over different density ranges, general theory of relativity and
possibly quantum chromodynamics  under extreme conditions. Astrophysical
observations of NS properties thus open the possibility of lending a
complementary vista to constrain the nuclear matter parameters (NMPs)
(characterizing the nuclear EoS) in sync with laboratory experiments.

The precise observations of high mass pulsars such as PSR J$1614-2230$
($1.908 \pm {0.016} ~M_\odot$)  \cite{Arzoumanian2018}, PSR
J$0348+0432$ ($2.01\pm0.04\,M_\odot$)  \cite{Antoniadis2013} and
the recently detected millisecond pulsar J$0740+6620$  
($2.14 {\scriptsize\begin{array}{c}+0.10\\-0.09\end{array}}M_\odot$)
\cite{Cromartie2019} have already put tight bounds on nuclear
matter EoS. Along with precise measurement of the NS masses, future
observations such as those planned by NICER (Neutron  star Interior
Composition Explorer) mission \cite{Keith2016,Arzoumanian2014},
eXTP (enhanced X-ray Timing and Polarimetry) Mission
\cite{Watts:2018iom}, LOFT (Large Observatory For X-ray Timing)
satellite \cite{Wilson-Hodge:2016grm}, and  ATHENA (Advanced  Telescope
for High  Energy Astrophysics) \cite{Motch:2013wfn}  may provide
besides the mass also the possible range for the  radius ($R_{1.4}$)
of a  canonical NS ($M  = 1.4M\odot$) and other selected NS. The
current empirical estimates of $R_{1.4}$  is $\simeq 11.9\pm 1.22$
km\cite{Bauswein18,Lim18,Most2018,Malik18,Radice2019}. Recently NICER came
up with one measurement of a radius $12.71_{-1.19}^{+1.14} \mathrm{km}$
for the NS with mass $1.34_{-0.16}^{+0.15} \mathrm{M}_{\odot}$
\cite{Miller2019,Riley2019}. However, more precise values of
masses and radii are required to impose stringent constraints on
the EoS.  Lately, after the detection of gravitational waves from
the GW170817 binary neutron star merger event \cite{Abbott17},
many authors looked into the rich connection between the  quadrupole
deformation and the very small nuclear objects more intensely
\cite{Radice2018,De18,Annala2018,Malik18,Fattoyev2018,Tews18}. The
gravitational wave phase evolution caused by that deformation can be
decoded by determining the dimensionless tidal deformability parameter
$\Lambda$ \cite{Flanagan2008,Hinderer2008, Hinderer2010,Damour2012}. It
is a measure of the response to the gravitational pull on the neutron
star surface correlating with the pressure gradients inside the NS and
strongly depends on the internal structure of the NS or on the EoS.
The future precise measurement of $\Lambda$ and radius of NS can be used
as an efficient probe on the  investigation of dense nuclear matter EoS.

The correlation systematics has proven to be a useful tool to
constrain the EoS,  thus the   NMPs which are its key ingredients
\cite{Vidana2009,Ducoin10,Ducoin11,Newton2011,Alam2016, Agrawal2020}.
Exploiting the thermodynamic Euler equation and the accepted broad
view of nuclear interaction, in a nonrelativistic framework, it has
been shown \cite{De2015} that an EoS for symmetric nuclear matter 
can be built up and that the thermodynamic state variables of nucleonic
matter (energy, pressure, incompressibility etc.) are coupled in a
correlated chain.  For given values of the energy per particle $e_0$ and the
nucleon effective mass $m_0^*$, all at saturation density $\rho_0$, the
direction of change in the incompressibility coefficient  ($K_0$) dictates
the direction of change in the skewness parameter ($Q_0$) in such a
way so as to keep $e_0$ invariant.  From application of different EoSs
intended to give the best fits to the diverse experimental data on a host
of finite nuclei, it is found that empirical values of $e_0$, $m_0^*$
and $\rho_0$ are obtained with so little scatter that the imprint of
the aforesaid correlation is still borne out.  For asymmetric nuclear
matter, similar observations are made that the symmetry energy
is correlated with higher order density derivatives \cite{Mondal2018}.
Correlations of different nuclear observables are also known to surface
out in different contexts; in the ambit of the droplet model, an
approximate analytical relation was found between $L_0$ and the neutron
skin thickness \cite{RocaMaza:2011pm} $\Delta r_{np}$ of asymmetric
nuclei. This lead to constraining $L_0$ when $\Delta r_{np}$ are taken
to be known from hadronic probes.

In the recent past, there have been several attempts to
constrain the behavior of EoSs from the tidal deformability
parameter using a diverse set of mean-field  models
\cite{Fattoyev2018,Malik18,Malik2019,Krastev2018}, which  satisfy
some basic properties of finite nuclei. Similar studies are also
carried out to constrain the EoS in a model independent manner
\cite{Annala2018,Zhang2018,Zhang2019,Carson2019a,Ferreira2019,Guven2020}.
In particular, the method of construction of nuclear meta-models
\cite{Margueron2018a,Margueron2018b} based on the Taylor expansion
around the saturation density $\rho_0$ has proved to be useful; the
expansion coefficients are identified with the NMPs. Experimental
values of the NMPs generate a model independent EoS; this further
enables one to study the effects of independent variation
of the NMPs on the properties of neutron stars with the allowance to
generate models that satisfy on average the constraints set on nuclear
matter properties at saturation. A recent result in this context draws
particular attention \cite{Carson2019a}. A regular set of Skyrme or
relativistic mean field (RMF) nuclear models \cite{Alam2016,Malik18}
fitted to nuclear properties include inherently the correlation among
the various NMPs \cite{Vidana2009,Ducoin11}. Whereas these EoSs show a
strong correlation of the NS radius or the tidal deformability of the
NS with the NMPs, inclusion of a diverse set of EoSs \cite{Carson2019a}
generated from independent variation of NMPs dilutes the correlation
casting doubt on the suitability of NS observables on constraining
the NMPs. The purpose of this communication is to identify the factors
which govern the correlations of tidal deformability with the symmetry
energy parameters.

In pursuance of our exploration, we employ a statistical chi-square based 
covariance approach (CCA) \cite{Reinhard2010} in the Skyrme framework to study the correlations of tidal deformability of neutron stars with the NMPs. This approach
enables one to study the correlations between a pair of quantities,
consistent with the fit data, with the help of Hessian matrix. In this
process the effects of the correlations among various NMPs imposed by
finite nuclei are inherently accounted through the Hessian matrix. We also
construct large number of EoSs using Multivariate Gaussian Distribution
(MVGD) by varying the NMPs independently as well as by including the
important correlations among them.  Comparison of these results with
those obtained within the CCA allows us to identify the
most important correlation among NMPs which helps in reconciling the
results from different investigations which are at variance otherwise.

{\it The EoS--}
The energy per nucleon, $e (\rho,\delta)$ for infinite asymmetric
nuclear matter in the Skyrme framework depends on the total
nucleonic density $\rho = (\rho_n + \rho_p)$ and the asymmetry parameter
$\delta=\frac{\rho_n -\rho_p}{ \rho} $ as,
\bea
\label{eq15}
e (\rho, \delta)&=& \frac{3}{5} \frac{\hbar^{2}}{2 m}
\left(\frac{3 \pi^{2}}{2}\right)^{2 / 3} \rho^{2 / 3} F_{5 / 3}+\frac{1}{8}
t_{0}
\left[2\left(x_{0}+2\right)\right.\nonumber\\
&&\left. -\left(2 x_{0}+1\right) F_{2}\right] \rho 
+ \frac{1}{48} t_{3} \rho^{\alpha +1 }\left[2\left(x_{3}+2\right) \right .\nonumber \\
&& \left. -\left(2 x_{3}+1\right) F_{2}\right] 
+\frac{3}{40}\left(\frac{3 \pi^{2}}{2}\right)^{2/ 3} \rho^{5/3}\nonumber \\ 
&& \left\{\left[\frac{2 \theta_s + \theta_{\rm sym}}{3}\right] F_{5 /
3} -\frac{1}{2}\left[\frac{\theta_s + 2 \theta_{\rm sym}}{3} \right] F_{8 / 3}\right\}\\
{\text{where}}&&
F_{m}(\delta)=\frac{1}{2}\left[(1+\delta)^{m}+(1-\delta)^{m}\right].
\nonumber 
\eea
The Skyrme parameters $t_0,t_3,x_0,x_3,\alpha, \theta_s$ and $\theta_{\rm sym}$
can be determined from the fit to the plethora of finite nuclear data;
expressions for the NMPs such as $e_0$,$\rho_0$,$K_0$, $Q_0$,
symmetry energy $J_0$, its density slope $L_0$ and the curvature parameter
$K_{\rm sym,0}$ can then be obtained. Conversely, from given values of
the NMPs, the seven Skyrme parameters mentioned can be uniquely
determined. The parameters $\theta_s$ and $\theta_{\rm sym}$ are measures of
the isoscalar and isovector nucleon effective masses, respectively. 
In this work, we represent the Skyrme EoS as given
by Eq. \ref{eq15} as a point in the seven dimensional space of NMPs,
$e_0,\rho_0, K_0, Q_0, J_0, L_0$, and $K_{\rm sym,0}$. Symbolically,
the $n$-th  EoS  in this space is written as
\bea
\text{EoS}_n^{\rm Skyrme} & =& 
\{{e_0},{\rho_0},{K_0},{Q_0},{J_0}, {L_0} 
{~\rm and~} {K_{\rm sym,0}}\}_n \nonumber \\  
&\sim &N(\boldsymbol{\mu},\boldsymbol{\Sigma}) \label{eos_mvgd}
\eea
where $N(\boldsymbol{\mu},\boldsymbol{\Sigma})$ is a
MVGD for NMPs  with  $\boldsymbol{\mu}$ being the
mean value of the nuclear matter parameters $\boldsymbol{p}$ and
$\boldsymbol{\Sigma}$ the covariance matrix. The diagonal elements of
$\boldsymbol{\Sigma}$ represent the variance or the squared error for
the ${p_i}$. The off-diagonal elements of $\boldsymbol{\Sigma}$ are
the covariance between different ${p_i}$  and yield the values of the
correlation  coefficients $r$ among  them.  Once, the $\boldsymbol{\mu}$
and $\boldsymbol{\Sigma}$ are known a large number of EoSs for the MVGD
of NMPs can be obtained.

{\it Estimation of $\boldsymbol{\mu}$, $\boldsymbol{\Sigma}$ and
$r$ --} 
The values of $\boldsymbol{\mu}$, $\boldsymbol{\Sigma}$ and
$r$ obtained for a single model within the covariance approach are
consistent with the fit data.  Where as, these quantities calculated
for a set of models yields  only  the model averages.  The quantity
$\boldsymbol{\mu}$ within the CCA corresponds  to the
values of NMPs obtained for the best fit parameters.  The covariance of
a pair of quantities $\mathcal{A}$ and $\mathcal{B}$ can be evaluated
within this  approach as,
\begin{equation}
\label{cov}
\boldsymbol{\Sigma}_{ \mathcal{A} \mathcal{B}}=
\sum_{\alpha \beta}\left(\frac{\partial \mathcal{A}}
{\partial \boldsymbol{q}_{\alpha}}\right)_{\boldsymbol {q}_{0}} \mathcal{C}_{\alpha \beta}^{-1}
\left(\frac{\partial {\mathcal{B}}}{\partial \boldsymbol {q}_{\beta}}\right)_{\boldsymbol {q}_{0}}
\end{equation}
{where $\boldsymbol {q}_\alpha$ and $\boldsymbol {q}_\beta$ are the
model parameters and  $\boldsymbol {q}_0$ represents the set of best
fit parameters \cite{Dobaczewski14,Reinhard2010}.  The quantities
$\mathcal{A}$ and $\mathcal{B}$ could be in general either the
NMPs, any observables, model parameters or even the mix quantities.
The $\mathcal{C}_{\alpha \beta}^{-1}$ is an element of the inverse of
the curvature or Hessian matrix given by},
\begin{equation}
\label{curv}
\mathcal{C}_{\alpha \beta}=\frac{1}{2}\left(\frac{\partial^{2} 
\chi^{2}(\boldsymbol {q})}{\partial \boldsymbol {q}_{\alpha}
 \partial \boldsymbol {q}_{\beta}}\right)_{\boldsymbol {q}_{0}}
\end{equation}
{with $\chi^{2}(\boldsymbol {q})$ being the merit function. The
covariance $\boldsymbol{\Sigma}_{\mathcal{A} \mathcal{B}}$ is thus
consistent with the set of fit data through the matrix $\mathcal{C}$. If
we consider $\mathcal{A,B}=\{e_0,\rho_0,K_0,Q_0,J_0,L_0,K_{\rm sym,0}\}$
which corresponds to a set of seven NMPs as discussed in above, a $7
\times 7$ covariance matrix can be obtained using Eq. \ref{cov}. Its
diagonal elements $ \boldsymbol{\Sigma}_{\mathcal{A A}}$ are the squared
errors and the off-diagonals elements $  \boldsymbol{\Sigma}_{\mathcal{A
B}}$ ($\mathcal{A} \neq \mathcal{B}$) are the covariance among them which
can be calculated using Eq. \ref{cov}.  The correlations among  the pair
of quantities $\mathcal{A}$ and $\mathcal{B}$ can be quantified using
the elements of covariance matrix as, }
\begin{equation}
\label{r_cor}
r_{\scriptstyle \mathcal{A B}}=
\frac{ \boldsymbol{\Sigma}_{\mathcal{A B}}} 
{\sqrt{ \boldsymbol{\Sigma}_{\mathcal{A A}} 
\boldsymbol{\Sigma}_{\mathcal{BB }}}}
\end{equation}
The absolute value of correlation coefficient $\mid r_{\scriptstyle
\mathcal{A B}} =1\mid $ indicates a perfect linear relation between
the quantities $\mathcal{A} $ and $\mathcal{ B} $. It is usually found that
if the correlations among various parameters are strong, then, the
errors associated with these individual parameters are also larger. In
other words, the correlated errors could be significantly larger than
uncorrelated ones \cite{Mondal2015}.  For example stronger correlations
among $K_0-Q_0$ and $L_0-K_{\rm sym,0}$ may result in larger errors
on $K_0$,$Q_0$,$L_0$ and $K_{\rm sym,0}$.  Further, propagation of
these errors affects other parameters.The errors on various parameters
and correlations among them can not be treated independently and
are driven by the fit data in the CCA. Nevertheless,
one also often calculates the  correlation coefficient $r_{\scriptstyle
\mathcal{A B}} $ for a set of models. In this case, the value of $
\boldsymbol{\Sigma}_{\mathcal{A B}}$ is given  as \cite{Brandt97},
\begin{equation}
\label{mod_cor}
\boldsymbol{\Sigma}_{\mathcal{A B}} = 
\frac{1}{N_m}\sum_i \mathcal{A}_i
\mathcal{B}_i -\left(\frac{1}{N_m}\sum_i \mathcal{A}_i\right)
\left(\frac{1}{N_m}\sum_j \mathcal{B}_j \right ),
\end{equation}
where the indices $i,j$ run over the number of models $N_m$.

{\it Results--}
\label{sec3}
The correlation between a pair of quantities within the CCA 
is calculated using the Hessian matrix which is
consistent with the set of fit data. We use this approach to study the
correlations of the tidal deformability with the slope and curvature
of nuclear symmetry energy and generate the confidence ellipses which
are consistent with the selected ground and excited state properties
of finite nuclei as well as with the maximum mass of NSs. The ground
state properties of finite nuclei considered are the binding energy
and charge radii. The excited state properties considered are the
isoscalar giant monopole resonance energy and the dipole polarizability.
Alternatively, the correlations of the tidal deformability with symmetry
energy parameters are studied using a set of EoSs obtained by varying
the NMPs independently.  In this process,  the correlations among the
various NMPs as imposed by the fit data are ignored.  In the following,
we will present our results obtained for independent and correlated
MVGD of NMPs  and compare them with the ones obtained using the  CCA.
\begin{figure}[htb]
 \centering
    \includegraphics[width=.45\textwidth,angle=0]{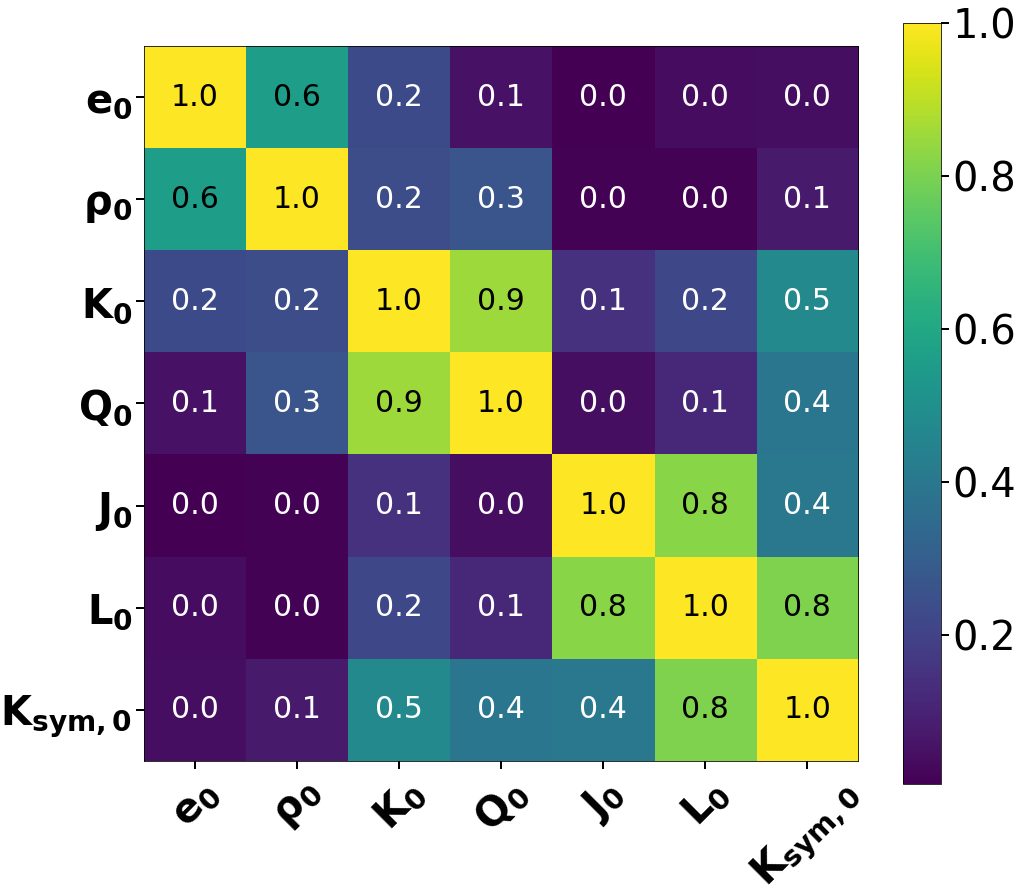} 
\caption{Correlations among various NMPs obtained using 237 selected
Skyrme models from Ref. \cite{Dutra2012}.  The correlations among the
off-diagonal pairs $K_0-Q_0$, $J_0-L_0$ and $L_0-K_{\rm sym,0}$ are
noticeable.} \label{fig1} \end{figure}

\begin{table}[]
  \begin{center}
  \centering
\setlength{\tabcolsep}{10pt}
\renewcommand{\arraystretch}{1.4}
    \caption{The mean value $\boldsymbol{\mu}_{p_i}$ and  error $\sqrt{
\boldsymbol{\Sigma}_{p_i p_i}}$ for the nuclear matter   parameters
${p_i}$ employed for the multivariate Gaussian distribution.  All the
quantities are in the units of MeV except for $\rho_0$ which is in unit
of fm$^{-3}$. We sample our EoSs  for three different cases; see text
for details.  The NMPs corresponding to  ${\rm Sk}\Lambda267$ model are
also listed.}
\label{tab1} \begin{tabular}{ccccc}
\hline \hline
 & \multicolumn{2}{c}{MVGD}         & \multicolumn{2}{c}{${\rm Sk}\Lambda267$}        \\ \cline{2-5} 
   ${p_i}$ & $\boldsymbol{\mu}_{p_i}$  & $\sqrt{\boldsymbol{\Sigma}_{p_i p_i}} $ &
   $\boldsymbol{\mu}_{p_i}$ & $\sqrt{ \boldsymbol{\Sigma}_{p_i p_i}} $ \\ \hline
$e_0$           & -16.0              & 0.25                 & -16.04             & 0.2                  \\
$\rho_0$        & 0.16               & 0.005                & 0.161              & 0.002                \\
$K_0$           & 230.0              & 20                   & 230.2             & 6.1                 \\
$Q_0$           & -300               & 100                  & -366.8            & 12.0                \\
$J_0$           & 32.0               & 3                    & 31.4              & 3.1                 \\
$L_0$           & 60.0               & 20                   & 41.1               & 18.2                \\
$K_{\rm sym,0}$ & -100.0             & 100                  & -124.0             & 70.2    \\ \hline \hline            
\end{tabular}
  \end{center}
\end{table}

Before embarking on our main results, we identify the important
correlations among the different NMPs using 237 Skyrme models
\cite{Dutra2012,Mondal2017}.  Most of these models are obtained by fitting
a few selected properties of finite nuclei that impose constraints
on NMPs'. In Figure \ref{fig1}, we present the $7 \times 7$ matrix  for
the correlation coefficients  obtained using Eqs. (\ref{r_cor})
and  (\ref{mod_cor}) for these 237 Skyrme models. The correlations among $K_0-Q_0$,$J_0-L_0$ and $L_0-K_{\rm
sym,0}$ pairs are noticeable. It may be emphasized that all the seven NMPs considered can be varied more or less independently  within  the Skyrme model. The correlations among the NMPs are the reflections of constraints
imposed by finite nuclei. The strong $L_0 - K_{\rm sym,0}$ correlation has also been observed earlier for other nuclear models
\cite{Chen2009,Danielewicz2009,Vidana2009,Ducoin11,Providencia2014}.

To elucidate the difference in results between the CCA and that based on
the MVGD of NMPs, We generate three different distributions of the NMPs,
namely, Case-I, Case-II and Case-III and obtain the corresponding sets of
the Skyrme EoSs. The mean values and the errors on each of the  NMPs are
exactly the same for the Cases I and II as listed in Table \ref{tab1}. The
Case I corresponds to the independent distribution of NMPs, i.e., the
correlation among different NMPs are ignored. In the Case II the $L_0-
K_{\rm sym,0}$ correlation is switched on and the correlation coefficient
is assumed to be $0.8$. We observe that the results for the tidal deformability of neutron stars are mainly sensitive to the $L_0-K_{\rm sym,0}$ correlations, thus, other correlations among NMPs as seen in Fig. \ref{fig1} are not considered here. The Case III is similar to Case II but the values of  $e_0$, $\rho_0$, $K_0$ and $Q_0$ are kept fixed to their mean values. The distributions of NMPs for all the three  Cases are filtered out such that the EoSs satisfy the causality  condition and yield the maximum mass of NS above 1.8 M$_\odot$. {The central value for the maximum NS mass for each of the distributions is $\sim 2.01~M_\odot$. The number of filtered EoSs for each of the distributions is about 3000.  {These three
distributions will allow us to unmask  how the existing correlations
among the NMPs may affect the correlation between the NS properties and
the NMPs, and how much the uncertainty on the NMPs will destroy possible
existing correlations. The Case III is considered in view of the
small uncertainties on the isoscalar nuclear matter  parameters obtained
within the CCA for a Skyrme model ${\rm Sk}\Lambda267$ \cite{Malik2019}
as listed in Table \ref{tab1}.  The fit data for this
model includes isoscalar and isovector giant resonances properties
of finite nuclei as discussed above together with the maximum NS mass.
The correlation coefficient among $L_0 - K_{\rm sym,0}$ is 0.9 for the ${\rm Sk}\Lambda267$ model. It may be noticed from Table \ref{tab1} that
the central values for the NMPs for the different Cases considered are
somewhat different from those for the ${\rm Sk}\Lambda{267}$ model. We
will see below that the trends of the results are mainly governed by
the uncertainties  and the correlations among different NMPs.

\begin{table}[!htb] \setlength{\tabcolsep}{12pt}
\caption{\label{tab2}The values of the
correlation coefficients for $\Lambda_{1.0,1.4,1.8}$ with $L_0$ and
$K_{\rm sym,0}$ for three different Cases considered and compared with
those for ${\rm Sk}\Lambda267$.} 
\renewcommand{\arraystretch}{1.0} \begin{tabular}{ccccc} \hline \hline
                 &      & $\Lambda_{1.0}$ & $\Lambda_{1.4}$ &
                 $\Lambda_{1.8}$  \\ \hline
\multirow{2}{*}{Case I} & $L_0$			& 0.82	  & 0.56	&
0.22    \\
		       & $K_{\rm sym,0}$	& 0.26	  & 0.58
		       &0.71	      	\\

\multirow{2}{*}{Case II} & $L_0$		& 0.9	  &0.83
&0.7		 	 \\
		       & $K_{\rm sym,0}$	& 0.84	  &0.86
		       &0.8			 \\

\multirow{2}{*}{Case III} & $L_0$		& 0.96	  &0.91
&0.82		 	  \\
		       & $K_{\rm sym,0}$	& 0.92	  &0.97
		       &0.98			 \\

\multirow{2}{*}{Sk$\Lambda$267} & $L_0$			& 0.92	  &0.85
&0.76		 	 \\
		       & $K_{\rm sym,0}$	& 0.89	  &0.94
		       &0.98		  \\
\hline \hline \end{tabular} 
\end{table}

\begin{figure*}[!htb]
 \centering
\begin{tabular}{c}
    \includegraphics[width=0.95\textwidth,angle=0]{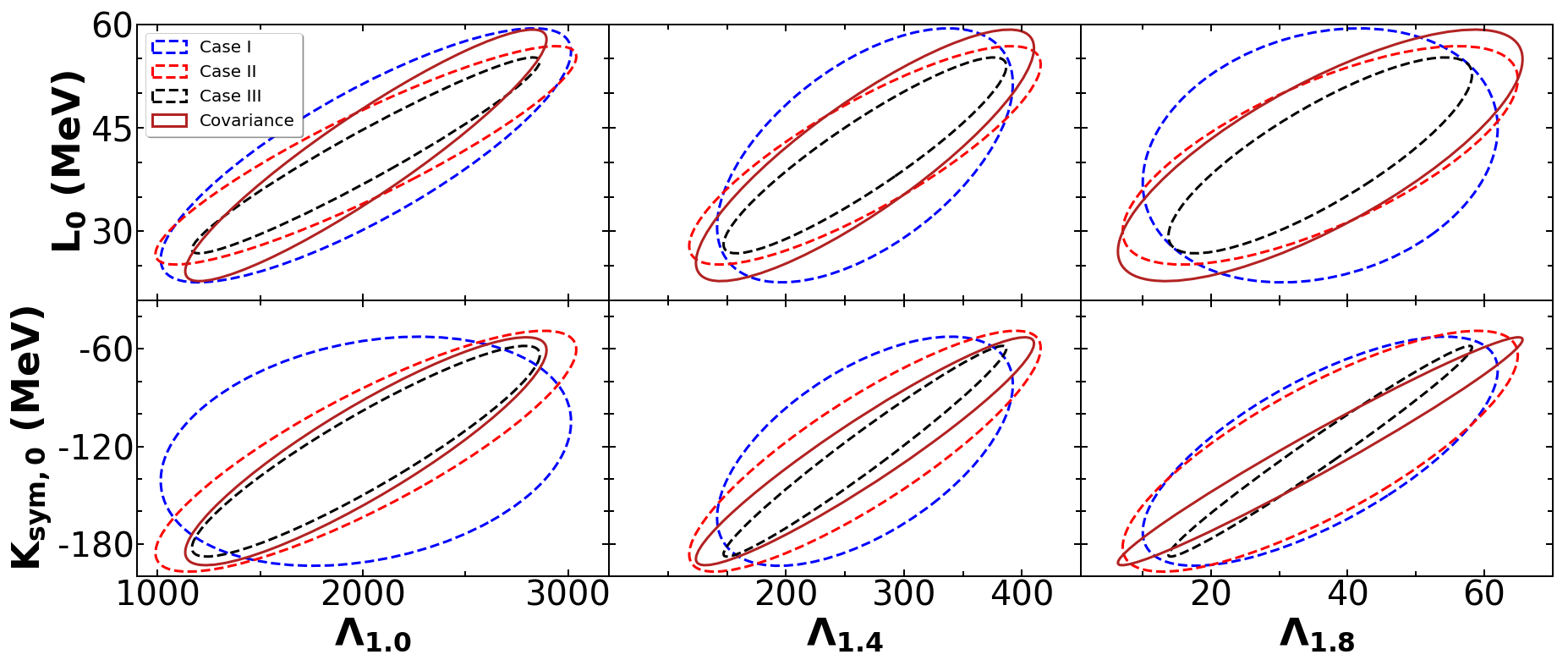} 
\end{tabular}
\caption{The $1\sigma$ confidence ellipses in the planes of
$\Lambda_{M}-L_0$ (top) and $\Lambda_{M}-K_{\rm sym,0}$ (bottom) with
$M=1.0,1.4$ and $1.8$  M$_\odot$ obtained for the Case I, II and III
and ${\rm Sk}\Lambda267$.  The central values of all the quantities
for all the Cases are matched to those for ${\rm Sk}\Lambda267$ for
the appropriate comparison. The actual central values for these Cases
are $L_0=60, K_{\rm sym,0}=-100, \Lambda_{1.0}=3200, \Lambda_{1.4}=430
~{\rm and}~ \Lambda_{1.8}=60$.} \label{fig2} \end{figure*}

In Figure \ref{fig2} we plot the confidence ellipses for
$\Lambda_{M}$ versus $L_0$ and $K_{\rm sym,0}$ for NS mass $M =1,1.4$
and 1.8 $M_\odot$ obtained for the Cases I, II and III and compare them with the
ones for the ${\rm Sk}\Lambda267$ model obtained within the CCA.  {The central values for $\Lambda_{M}$, $L_0$ and $K_{\rm
sym,0}$ for Cases I, II and III are matched to those for the ${\rm
Sk}\Lambda267$ model for the appropriate comparison.} {Through
this comparisons, we would like to identify the reasons which can
be attributed to the marked differences in the correlations of the
tidal deformability with $L_0$ and $K_{\rm sym,0}$     as reported
earlier \cite{Ferreira2019, Carson2019a, Fattoyev2018}.  }
The values of correlation coefficients for the results presented in the
figure are summarized  in Table \ref{tab2}.
Sometimes, the calculations are performed by ignoring the correlation
among NMPs, which is analogous to our Case I. This yields weak
correlations as seen earlier \cite{Carson2019a}.  {For instance,
correlations of $\Lambda_{1.0,1.4,1.8}$ with $K_{\rm sym,0}$
are $ r \sim 0.3$ - 0.7.} The narrowing of the confidence ellipses
for  Case II indicate  stronger correlations of $\Lambda_{1.0}$ and
$\Lambda_{1.4}$ with $L_0$ and $K_{\rm sym,0}$,  $ r \sim 0.8$ - 0.9,
while, these correlations become moderate for $\Lambda_{1.8}$. The
${\rm Sk}\Lambda267$ model predicts stronger
correlations for $\Lambda_{M}-L_0$ and $\Lambda_{M}-K_{\rm sym,0}$
pairs for all NS masses considered. The $\Lambda_{M}- L_0$
correlations decrease only marginally with increasing mass of NS and
$\Lambda_{M}-K_{\rm sym,0}$ correlations show the opposite trend.
Upon examining closely we notice that the uncertainties obtained
for the isoscalar NMPs such as $e_0$, $\rho_0$, $K_0$ and $Q_0$ for the
${\rm Sk}\Lambda267$ model are  smaller than those employed for the Case I
and II (see Table \ref{tab1}). Thus it appears that the larger uncertainties
on $e_0$, $\rho_0$, $K_0$ and $Q_0$ are also responsible for masking
or  reducing the correlations of $\Lambda_{1.8}$ with $L_0$ and $K_{\rm
sym,0}$ for the Cases I and II.  These observations are reinforced in
Case III by freezing the values of the isoscalar NMPs to their central
values. The results for Case III are in qualitative agreement with those for
${\rm Sk}\Lambda267$. These results emphasize that the correlation of
the tidal deformability with $L_0$ and $K_{\rm sym,0}$ are sensitive to
the distributions of the NMPs used. 
In particular, the correlations are weaker if the
NMPs are independently varied.  The distribution of NMPs should be
consistent with the finite nuclei data and the other relevant observables.
The interplay of isoscalar NMPs such $K_0$ and $Q_0$ in masking the
correlations of $\Lambda_M$ with $L_0$ and $K_{\rm sym, 0}$ can be
qualitatively understood from the Taylor expansion of EoS, 
\bea
\tilde{e}(\rho,\delta) &\approx&
  e_0 + J_0 \delta^2 +L_0 \epsilon\delta^2 + \frac{1}{2} (K_0 + K_{\rm sym, 0}
\delta^2 )  \epsilon^2 \nonumber \\
&& +\frac{1}{6}(Q_0 +  Q_{\rm sym, 0} \delta^2 ) \epsilon^3
+ ...
\eea
where,  $\epsilon = \frac{\rho -\rho_0}{3\rho_0}$.  It is evident from
the above equation  that for the pure neutron matter (i.e. $\delta =
1$), the isoscalar parameters $K_0$ is tangled with $K_{\rm sym, 0}$ so
that if $K_0$ increases, $K_{\rm sym, 0}$ decreases and vice versa; so is
the case with $Q_0$ and $Q_{\rm sym, 0}$.  For the $\beta$-equilibrated
matter, the asymmetry $\delta$ usually decreases with the density,
since, the proton fraction increases with density \cite{Burgio2018}.
For the neutron stars with mass $\sim 1M_\odot$, $\epsilon < 1, \delta
\approx 1.0$ at the center, the properties of the stars are predominantly
governed by $L_0$. On the other hand, near the maximum mass $M_{max}\sim
2.0M_\odot$, which corresponds to the central densities having $\epsilon
\gtrsim 1, \delta \approx 0.4 - 0.8$, their properties are governed by
$K_0$, $Q_0$. At the intermediate masses, both the isoscalar and isovector
NMPs  play an important role in determining the properties of neutron stars
\cite{Alam2016}.  Since, the values of $e_0$, $\rho_0$, $K_0$ and $Q_0$
are tightly determined by our fit data, it facilitates in identifying
the correlations of the  $\Lambda$ with $L_0$ and $K_{\rm sym, 0}$.

\begin{figure}
 \centering
\begin{tabular}{c}
    \includegraphics[width=.48\textwidth,angle=0]{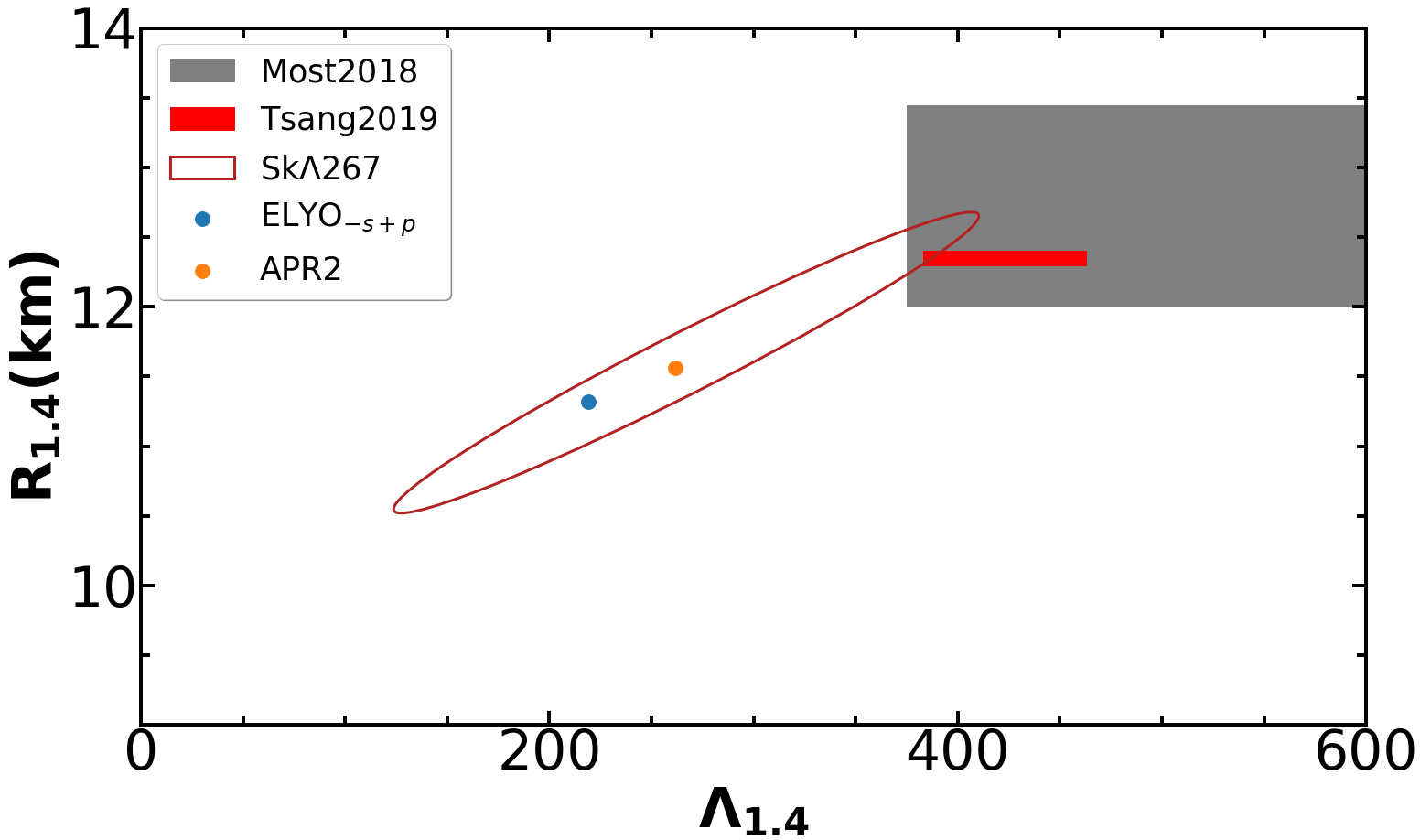}
\end{tabular}
\caption{The $1\sigma$ confidence ellipse in the plane of
$\Lambda_{1.4}-R_{1.4}$ for ${\rm Sk}\Lambda267$ model compared with  some
recent results  \cite{Most2018,Tsang2019,Bonnard2020,Sabatucci2020}.} 
\label{fig3} \end{figure}

In Figure \ref{fig3}, we plot the variation of $\Lambda_{1.4}$
with $R_{1.4}$. The brown ellipse is the bound obtained from
${\rm Sk}\Lambda267$ model within the $1\sigma$ confidence.  {This
bound has good overlap with the recent result obtained with APR2
\cite{Sabatucci2020} and ELYO$_{-s+p}$ \cite{Bonnard2020} and agrees
marginally with the ones obtained by Most et al  \cite{Most2018} and
Tsang et al \cite{Tsang2019}.} It may be pointed out that the results
for Most {\it et al.} is within $2\sigma$ limits and obtained without
considering bounds from microscopic finite nuclei experiments.

{\it Conclusions--} \label{sec4} The  chi-square based covariance approach
has been applied to evaluate the correlations of the tidal deformability
$\Lambda$ of neutron stars with the slope $(L_0)$ and curvature $(K_{\rm
sym,0})$ of the nuclear symmetry energy.  This approach enables one to
calculate correlation among any desired observables for a given set of
fit data  without a prior  knowledge of the distribution of the nuclear
matter  parameters.  The correlated distribution of NMPs in turn can be
constructed in this approach in a consistent manner through the  Hessian
matrix which implicitly depends on the fit data.  The correlations
of $\Lambda$  with $L_0$ and $K_{\rm sym,0}$ are also evaluated
by  employing explicitly a multivariate Gaussian distribution of NMPs
corresponding to commonly used values for their mean and variances with
and without inclusion of $L_0-K_{\rm sym,0}$ correlation. 
{Comparison of the results from the multivariate Gaussian distribution 
with those from the chi-square based covariance approach indicates that, 
in order to study the correlation systematics involving tidal deformability 
of neutron stars, the use of multivariate distribution of the nuclear matter parameters must be appropriately guided by a realistic and as complete as possible set of fit data.}
Employing a set of EoS which
corresponds to uncorrelated distribution of NMPs or even a correlated
distribution inconsistent with fit data may mask realistic correlations
that constrained nuclear models would be able to identify. The narrowing
down of the difference in distributions of NMPs between Case III where
the isoscalar uncertainties are frozen and that for ${\rm Sk}\Lambda267$
where the isoscalar uncertainties are small also possibly points out
to the importance of an interplay between the isoscalar and isovector
NMPs. The role of the distribution of NMPs on the correlation systematics
and their sensitivity to various fit data need to be further investigated
within the Bayesian analysis to unveil further the information content of the tidal deformability.

{\it Acknowledgement--} {We would like to acknowledge Chiranjib Mondal and
Bharat Kumar for the discussions and their suggestions. C.P. acknowledges
financial support by Fundação para a Ciência e Tecnologia (FCT)
Portugal under projects UID/FIS/04564/2019, UID/FIS/04564/2020,
and POCI-01-0145-FEDER-029912. J.N.D. acknowledges support from the
Department of Science and Technology, Government of India  with grant
no. EMR/2016/001512.}

%
\end{document}